\documentclass{PoS}

\title{Data Acquisition with GPUs: The DAQ for the Muon $g$-$2$ Experiment at Fermilab}

\ShortTitle{Data Acquisition with GPUs}

\author{\speaker{W.~Gohn}\thanks{for the Muon g-2 collaboration.}\\
        University of Kentucky\\
        E-mail: \email{gohn@pa.uky.edu}}

\abstract{Graphical Processing Units (GPUs) have recently become a valuable computing tool for the acquisition of data at high rates and for a relatively low cost. The devices work by parallelizing the code into thousands of threads, each executing a simple process, such as identifying pulses from a waveform digitizer. The CUDA programming library can be used to effectively write code to parallelize such tasks on Nvidia GPUs, providing a significant upgrade in performance over CPU based acquisition systems.

The muon $g$-$2$ experiment at Fermilab is heavily relying on GPUs to process its data. The data acquisition system for this experiment must have the ability to create deadtime-free records from 700 $\mu$s muon spills at a raw data rate 18 GB per second. Data will be collected using 1296 channels of $\mu$TCA-based 800 MSPS, 12 bit waveform digitizers and processed in a layered array of networked commodity processors with 24 GPUs working in parallel to perform a fast recording of the muon decays during the spill. The described data acquisition system is currently being constructed, and will be fully operational before the start of the experiment in 2017.}

\FullConference{38th International Conference on High Energy Physics\\
		3-10 August 2016\\
		Chicago, USA}

\begin{document}

\section{GPUs in Data Acquisition}

Graphical Processing Units (GPUs) are a useful tool for the processing of data in a moderate size high energy physics data acquisition system, as they have the potential to dramatically improve performance at a reasonable cost by parallelizing data processing. There are other technologies that can achieve an equivalent level of parallelization, such as FPGAs, but they are more complicated to program, and not as well supported. Because the GPU technology was developed initially for commercial gaming applications, there are a multitude of resources available for users to learn basic programming environments and maintenance technologies.  GPUs were the key that made the data acquisition scheme for the Muon g-2 experiment possible.

The need for parallelization of data processing is a common desire for HEP experiments, and there are many different technologies available to do so. In addition to numerous models of GPUs, there are also coprocessor systems such as the Intel Xeon Phi, which utilize fewer but faster cores than the GPUs, as well as FPGAs or ASDQs, which require significantly more programming overhead than do the GPUs systems. There are situations when these other types of systems will be more efficient, in particular for very large experiments or for very small and repetitive processes, but GPUs have a sweet spot  in the phase space of complexity and processing power in which they are the most effective solution.

The choice of a particular GPU or coprocessor architecture will of course depend on the specific application. GPUs have a large number of cores that run at a relatively low rate, and the Xeon Phi has fewer cores with a slight improvement of processor speed. There are many available GPU architectures, but a main point of difference is between GPUs that are developed for scientific applications and those developed for gaming and advances graphics. Gaming GPUs can have as many cores with equivalent specifications to scientific GPUs, at a fraction of the price. The main difference is that the scientific GPUs support ECC memory protection, which is highly desirable in a data acquisition system. Table~\ref{tab:arch} shows a comparison of a representative example of each type.

\begin{table}
\centering
\begin{tabular}{|c|c|c|}
\hline
Scientific GPU & Gaming GPU & Coprocessor \\
\hline
Nvidia Tesla K40 & Nvidia Black Titan & Intel Xeon Phi\\
2880 CUDA cores & 2880 CUDA cores & 68 Cores / 272 threads\\
740 MHz & 980 MHz & 1.4 GHz\\
12 GB memory & 6 GB memory & 384 GB memory\\
\hline
\end{tabular}
\label{tab:arch}
\caption{A representative example of different GPU varieties and a coprocessor system.}
\end{table} 

Peripheral computer components are also important for the operation of GPUs with high rates of data throughput. It is necessary to select a motherboard with the appropriate number of available PCIe slots, and if data rates are high, PCIe version 3.0 will nearly double the available data throughput versus PCIe version 2.0. It is also valuable for the motherboard to possess a PLX chip, which allows it to distribute the load over 40 lanes in order to operate all PCIe slots simultaneously at 16x.

Choices for programming the GPU algorithms depend largely on the GPU architecture selected. If an Nvidia brand GPU is used, the proprietary CUDA platform can be used for GPU programming. CUDA provides a set of libraries for C++ with compilation and profiling tools that can be used to easily develop and compile kernel algorithms to run on the GPU and fully and efficiently utilizes the multithreading capabilities. An alternative to CUDA is OpenCL, which has the advantages of being open source and working on many different architectures. For the Muon g-2 data acquisition development, we are using CUDA.

The use of GPUs in the data acquisition systems of particle physics experiments is becoming increasingly popular, and the next generation of experiments are using them heavily. GPUs are used for, or plan to be used for, online tracking of charged particle tracks in $\bar P$ANDA at FAIR~\cite{bianchi}, reconstructing ring-shaped hit patterns in a $\breve{C}$erenkov detector in NA62 at CERN~\cite{ammendola}, track and vertex reconstruction for Mu3e at PSI~\cite{vombruch}, and track pattern recognition in LHCb~\cite{gallorini}. For muon g-2, GPUs are used to identify islands of data in long waveform digitizer traces~\cite{gohn}.

\section{Motivation: Muon g-2}

The Muon g-2 experiment will begin commissioning at Fermilab in 2017. The experiment is designed to measure the anomalous magnetic moment of the muon $a_\mu$ to 140 ppb, which is 4 times better than the previous best measurement~\cite{bnl}. The 14 m diameter superconducting magnetic ring that was used in the BNL measurement was moved to Fermilab and installed in 2015, and the measurement and shimming of the magnetic field was completed in August 2016. Detector systems and electronics will be installed in late 2016, and we plan for initial beam in the hall in late Spring of 2017. 

For a more complete description of the experiment, please refer to articles in these proceedings by C. Polly, E. Swanson (overview), J. Kaspar (detectors), V. Tischenko (muon storage), B. Kiburg and D. Flay (megnetic field measurement), D. Sweigart (electronics), R. Bjorkquist (beam measurements), A. Epps, J. Mott, and T. Stuttard (straw trackers).

\section{The Muon g-2 DAQ}

The data acquisition system for Muon g-2 has been designed to utilize GPUs to extract the necessary information for the measurement of the muon precession frequency from digitized muon fills. The design consists of 26 Nvidia Tesla K40 GPUs housed by pairs in 13 front-end computers. Each front-end computer processes data from two calorimeters, with a dedicated GPU for each of the calorimeters. Construction of the data acquisition system is complete, and commissioning is underway.

Muons will be injected into the ring at an average fill rate of 12 Hz consisting of sequences of eight successive fills with 700 $\mu$s data collection periods separated by 10 ms. Approximately 16,000 muons are stored per fill. Positrons resulting from the muon decays are detected using twenty-four electromagnetic calorimeters, each comprised of 9$\times$6 arrays of PbF$_2$ crystals read out by SiPMs. The full 700 $\mu$s long waveform is stored by from each of the 1296 custom $\mu$TCA 800 MSPS, 12-bit waveform digitizers~\cite{sweigart}. For a 12 Hz spill rate the time-averaged rate of raw ADC samples is 18.6 GB/s in total.

\begin{figure}\centering
\includegraphics[width=8cm]{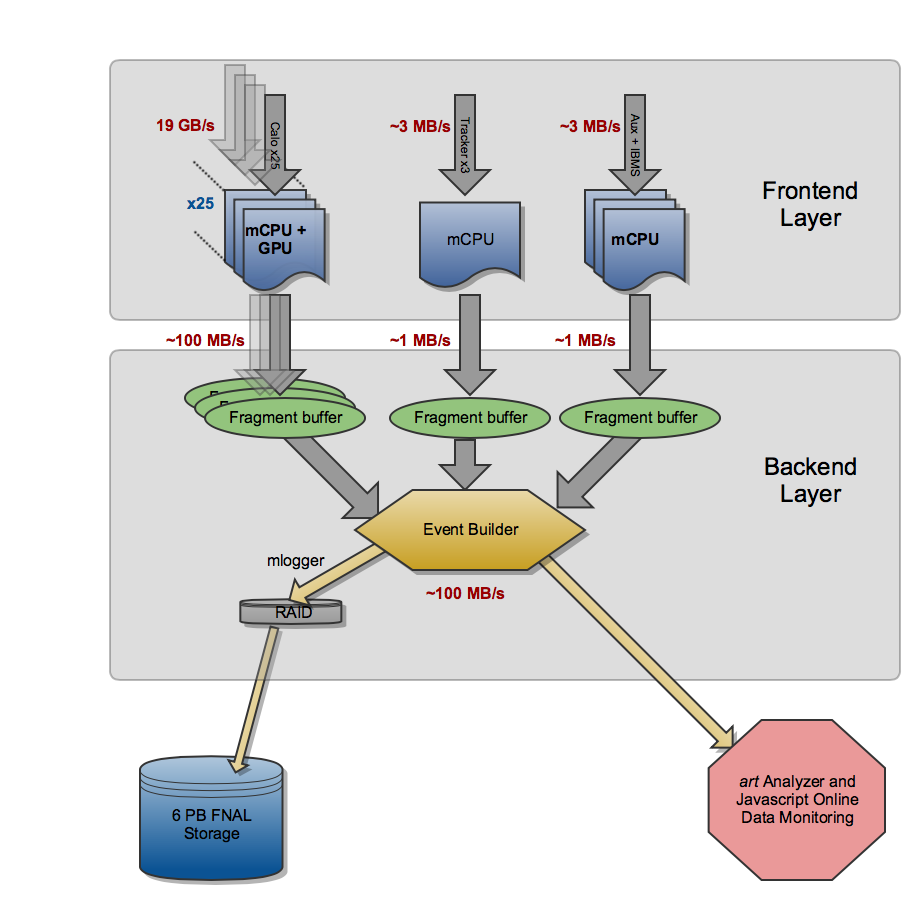}
\caption{A schematic of the data acquisition system for Muon g-2}
\label{fig:diagram}
\end{figure}

The DAQ is constructed using a layered array of commodity, networked processors, as shown in Fig.~\ref{fig:diagram}. The frontend layer reads out data from the digitizers and MHTDCs, unpacked and processes the data, and sends it the the backend layers that is used for assembly of event fragments and writing the data files to disk. A slow control layer is used for the setting and monitoring of high voltages, temperatures, and other slowly varying parameters. An online analysis layer using \emph{art}~\cite{art} and javascript monitors the integrity of raw and processed data.

The software for the Muon g-2 DAQ is based on the MIDAS data acquisition software~\cite{midas} developed at PSI and TRIUMF. The package includes a web interface for easy experiment control and the necessary machinery for acquiring data and writing it to disk. The frontend code is being written in C++ with CUDA, and we have created a javascript based online data quality monitoring system. Data will be written to disk as MIDAS data files.

The frontend code is multithreaded with mutual exclusion (mutex) locks, which insure data integrity. Three threads work in parallel to process the data. The first is the "tcp thread", which reads and unpacks data from a TCP socket and copies it to a ring buffer. Next a "GPU thread" performs a memcpy of the data from one fill to the GPU, where it is processed and returned as chopped islands. The GPU processing uses eight kernel routines that process each 560,000 sample waveform using a total of over 32 million threads, as shown in Table.~\ref{tab:kernel}. A third "MFE thread" then packs the data into MIDAS banks and performs a lossless compression algorithm before sending the data to the event builder.

\begin{table}
\begin{tabular}{|c|c|c|}
\hline
Function&	 Number of threads	& Execution time (ms)\\
\hline
Compute pedestals as average of first 100 samples&	54	&	0.1	\\
Determine if threshold is passed for each sample & 560k & 1.7\\
Add pre-samples and post-samples to each island& 560k	& 0.1	\\
Check to see if any islands have merged &	560k	&	0.2	\\
Save an array of identified islands &560k & 0.2\\
Sum all waveforms & 560k & 1.2\\
Decimate the sum for the Q-method & 17.5k & 0.3\\
Make a fill-by-fill sum of waveforms & 30M & 2.4\\
\hline
\end{tabular}
\caption{CUDA kernel routines used to process each muon fill}
\label{tab:kernel}
\end{table}

The GPU kernel functions are designed to perform two different algorithms on each waveform, which produce two independent output streams to be analyzed for measurements of the precession frequency. The two algorithms are called the T-method (time) and Q-method (charge). 

For the T-method, positron events in each calorimeter are individually identified, sorted and fit to obtain a spectrum of time and energy. All events above a certain threshold are included, and the precession frequency is determined from a fit to a pileup-subtracted histogram. This was the analysis method used in the Brookhaven experiment. To implement this in the GPU requires three individual kernel functions. The first determines if the threshold has been passed for each of the 560,000 samples (processing with one thread per sample). The second then adds a set number of postsamples and presamples to each island. The third checks if any islands are now overlapping, and combines them to form single larger islands. An array of island data is then passed out of the GPU and back to the computer memory. 

The second technique, the Q-method, does not identify individual positron events, but instead integrates the detector current as a proxy for the event energy. In practice this is performed by histogramming the full pedestal subtracted waveforms on a fill-by-fill basis, and fitting that sum to extract the frequency. 

The DAQ installation is nearly complete. A GPU cluster containing 28 (including spares) Nvidia Tesla K40 GPUs is up and running, and three of four required backend computers are operational. A 40 TB RAID has been installed for short-term data storage, and a fiber-optic 10 GbE network has been installed. A Meinberg GPS unit is used to timestamp fills as they come in to the DAQ.

\begin{figure}\centering
\includegraphics[width=12cm]{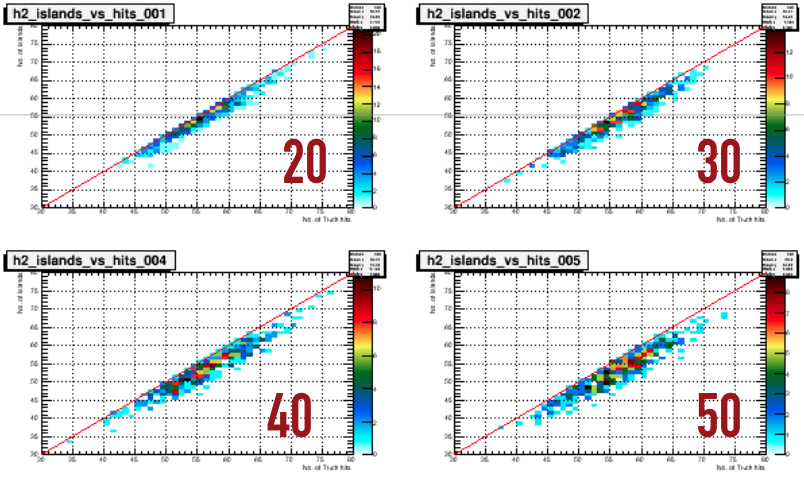}
\caption{An analysis of simulated data that has been processed by the DAQ. The vertical axis of each plot shows the number of truth islands from the simulation, and the horizontal axis is the number of identified islands. The red line is a visual aid to show a slope of 1, which is the ideal detection scenario. Each pannel shows the identified islands over a dataset using a different threshold value, written in red.}
\label{fig:sim}
\end{figure}

The testing and commissioning of the DAQ is ongoing. It has been tested using a simulator that generates realistic data rates for all 24 calorimeters, which allows for the testing of the GPU pulse-finding algorithms, as shown in Fig.~\ref{fig:sim}. The full DAQ was also used to acquire data from a full calorimeter at a test beam at SLAC in June, 2016. Characterization of the 1296 channels of WFDs required for the experiment is underway, and a subset of those are being read out using the full data acquisition system, and expansion to all 1296 is imminent. Full commissioning with all detector systems will take place at the beginning of 2017.

\section{Conclusion}

The data acquisition system for the Muon g-2 experiment has been constructed to heavily rely on GPUs for processing of the data. A small farm of 26 Nvidia Tesla K40 GPUs has been assembled and tested, and commissioning is underway. The use of GPUs in the DAQ will allow for a reduction of 20 GP/s of raw data into approximately 100 MB/s of processed data, that will be written to disk. The GPU is an important component and an enabling technology for this experiment.

\end{document}